\documentclass[a4paper]{spie}  

\usepackage[]{graphicx}

\title{Active and passive shielding design optimization and technical solutions for deep sensitivity hard
X-ray focusing telescopes} 

\author{G. Malaguti\supit{a}, G. Pareschi\supit{b}, P. Ferrando\supit{c}, E. Caroli\supit{a}, G. Di Cocco\supit{a}, L. Foschini\supit{a}, S. Basso\supit{b}, S. Del
Sordo\supit{d}, F. Fiore\supit{e}, A. Bonati\supit{f}, G. Lesci\supit{f}, J.M. Poulsen\supit{f}, F. Monzani\supit{f}, A. Stevoli\supit{f}, B.
Negri\supit{g}
\skiplinehalf
\supit{a} INAF, Istituto di Astrofisica Spaziale e Fisica Cosmica, Sezione di Bologna, via P. Gobetti 101, 40129 Bologna, Italy\\
\supit{b} INAF, Osservatorio Astronomico di Brera, via E. Bianchi 46, 23807 Merate Italy \\
\supit{c} DSM/DAPNIA/Service d'Astrophysique, CEA/Saclay, 91191 Gif-sur-Yvette Cedex, France \\
\supit{d} INAF, Istituto di Astrofisica Spaziale e Fisica Cosmica, Sezione di Palermo, via La Malfa 153, 90146 Palermo, Italy \\
\supit{e} INAF, Osservatorio Astronomico di Roma, via Frascati 33, 00040 Monteporzio, Italy\\
\supit{f} Alenia Spazio SpA LABEN, Strada Padana Superiore 290, 20090 Vimodrone, Italy \\
\supit{g} ASI, Agenzia Spaziale Italiana, Viale Liegi 26, 00198 Roma, Italy 
}

\authorinfo{Send correspondence to G. Malaguti, E-mail: malaguti@bo.iasf.cnr.it, Telephone: +39 051 6398679}

\begin{document} 
  \maketitle 

\begin{abstract}
The 10--100 keV region of the electromagnetic spectrum contains the potential for a dramatic
improvement in our understanding of a number of key problems in high energy astrophysics. A deep
inspection of the universe in this band is on the other hand still lacking because of the demanding
sensitivity (fraction of $\mu$Crab in the 20--40 keV for 1 Ms integration time) and imaging ($\approx15''$
angular resolution) requirements. The mission ideas currently being proposed are based on
long focal length, grazing incidence, multi-layer optics, coupled with focal plane detectors with
few hundreds $\mu$m spatial resolution capability. The required large focal lengths, ranging
between 8 and 50 m, can be realized by means of extendable optical benches (as foreseen e.g. for the
HEXIT-SAT, NEXT and NuSTAR missions) or formation flight scenarios (e.g. Simbol-X and XEUS). While
the final telescope design will require a detailed trade-off analysis between all the relevant
parameters (focal length, plate scale value, angular resolution, field of view, detector size, and
sensitivity degradation due to detector dead area and telescope vignetting), extreme attention must
be dedicated to the background minimization. In this respect, key issues are represented by the
passive baffling system, which in case of large focal lengths requires particular design
assessments, and by the active/passive shielding geometries and materials. In this work, the result
of a study of the expected background for a hard X-ray telescope is presented, and its implication
on the required sensitivity, together with the possible implementation design concepts for active
and passive shielding in the framework of future satellite missions, are discussed.
\end{abstract}


\section{INTRODUCTION}

The study of the universe above 10--20 keV is still hampered by the absence, up to now, of focusing telescopes for this energy band.
This region of the electromagnetic spectrum contains the potential for a dramatic improvement in our understanding of a number of key
astrophysical problems which remain still open, such as the origin of the cosmic X-ray background (CXB) in the 20--40 keV where its
energy density peaks, or the history of super-massive black holes (SMBH) growth$^1$. Despite its scientific importance, technological
problems have prevented so far the development of suitable hard X--ray telescopes.

The most sensitive experiment flown at these energies, BeppoSAX/PDS$^2$, had a flux limit of $\approx300\;\mu$Crab at $\approx$30
keV. However, because of its large field of view (FOV$\sim1.3^\circ$ in diameter) and lack of imaging capabilities, the experiment
was limited by high intrinsic background and source confusion. The IBIS imager$^3$ currently operative onboard the INTEGRAL
satellite, which is based on a coded aperture mask is, on the other hand, capable of only moderate imaging power (angular resolution
$\simeq12'$). Its large FOV ($\sim30^\circ\times30^\circ$ at zero response) however, determines a high background level which
results in a sensitivity slightly worse than the BeppoSAX/PDS one.

Recent technological advancements in the field of both X-ray mirrors and focal plane detectors, allow for the first time the
development of fine imaging, deep sensitivity high energy satellite-borne telescopes for the energy band above 10 keV, some of which
have already had a protoflight balloon test$^4$.  The use of X-ray focusing mirrors very significantly reduces the instrumental
background by concentrating the flux into a tiny region (few hundreds of $\mu$m) of the detector. However, since many of the key
science objectives require very long observations ($10^5\div10^6$ s), the residual focal spot area background can eventually
dominate the telescope's sensitivity. For these reasons, the evaluation of the expected background and its characterization in terms
of cosmic diffuse against particle induced component acquire paramount importance.

At present, different design philosophies have been proposed for the forthcoming hard X--ray focusing telescopes, which are based either on free flyer (HEXIT-SAT$^1$) or formation flight (Simbol-X$^5$, XEUS$^6$) mission scenarios. Since the in-flight background is expected to vary according to the mission technical design and orbit choices, dedicated studies are necessary. The present work is aimed at the study of the impact on the limiting sensitivity caused by the background and the other telescope key parameters. The science-driven design concepts and their possible implementations of passive and active shielding in the framework of the forthcoming satellite missions are then presented and discussed.

\section{TELESCOPE SENSITIVITY}

The minimum detectable flux for a focusing X-ray telescope is given by

\begin{equation}
\label{eqn:Fmin}
F_{\rm min} = n_{\sigma}\frac{\sqrt{B \cdot A_{\rm d} } }{ \varepsilon \cdot \eta \cdot (1-\beta) \cdot A_{\rm eff}^{i} \cdot \sqrt{N \cdot T \cdot \Delta E}}
\;\;\;\;\; {\rm photons\;cm^{-2}\;s^{-1}\;keV^{-1}}\;,
\end{equation}
where $B$ is the background flux (in counts cm$^{-2}$ s$^{-1}$ keV$^{-1}$), $A_{\rm d}$ is the dimension of the spot area of each
mirror module (cm$^{2}$), $A_{\rm eff}^{i}$ is the collection area of a single module weighted by the total mirror reflectivity
(cm$^{2}$), $N$ is the number of mirror modules, $\eta$ is the fraction of incoming X-ray photons reflected within $A_{\rm d}$,
$\varepsilon$ is the detector quantum efficiency, $\beta$ is the fraction of detector dead area (due to the pitch between two
adjacent pixels, plus the possible vignetting caused by collimator walls), $n_\sigma$ is the statistical significance of the
detection, $T$ the integration time (in s), and $\Delta E$ the energy band (in keV).

   \begin{figure}
   \begin{center}
   \begin{tabular}{c}
   \includegraphics[scale=0.8]{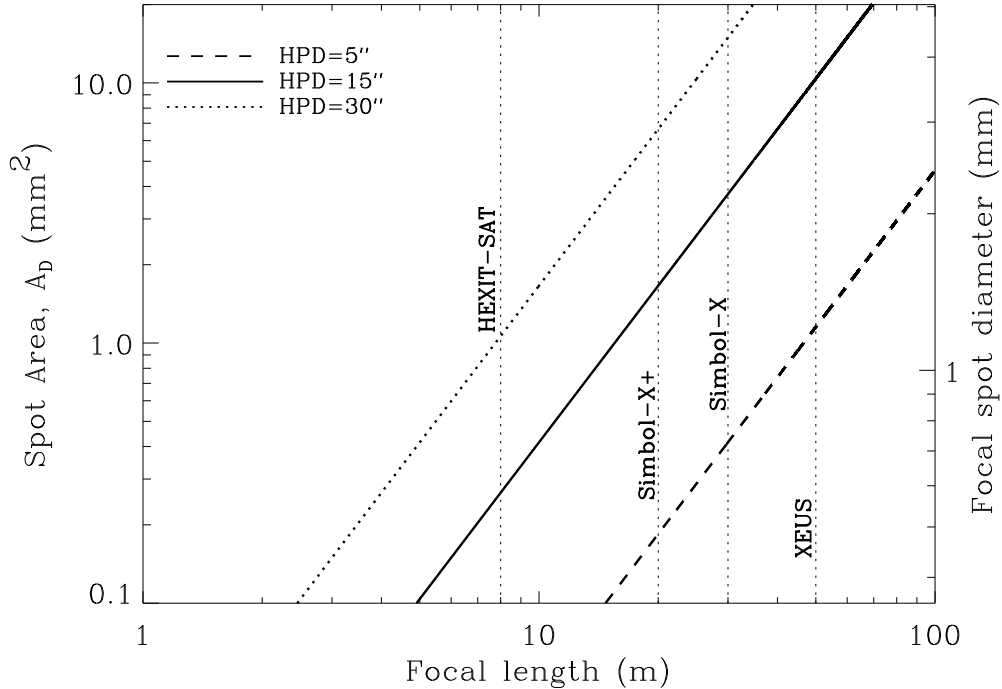}
   \end{tabular}
   \end{center}
   \caption[example] 
   { \label{fig:Ad_2} Telescope spot area $A_{\rm D}$ as a function of the focal length FL for different mission concepts and angular
resolutions HPD. The left hand $Y$ axis gives $A_{\rm D}$ in mm$^{2}$ to get a direct evaluation of the associated background variation.
The right hand $Y$ axis gives the linear dimension of the focal spot, for a direct comparison with the required detector spatial
resolution.}
   \end{figure} 

\begin{figure}
   \begin{center}
   \begin{tabular}{c}
   \includegraphics[scale=0.7]{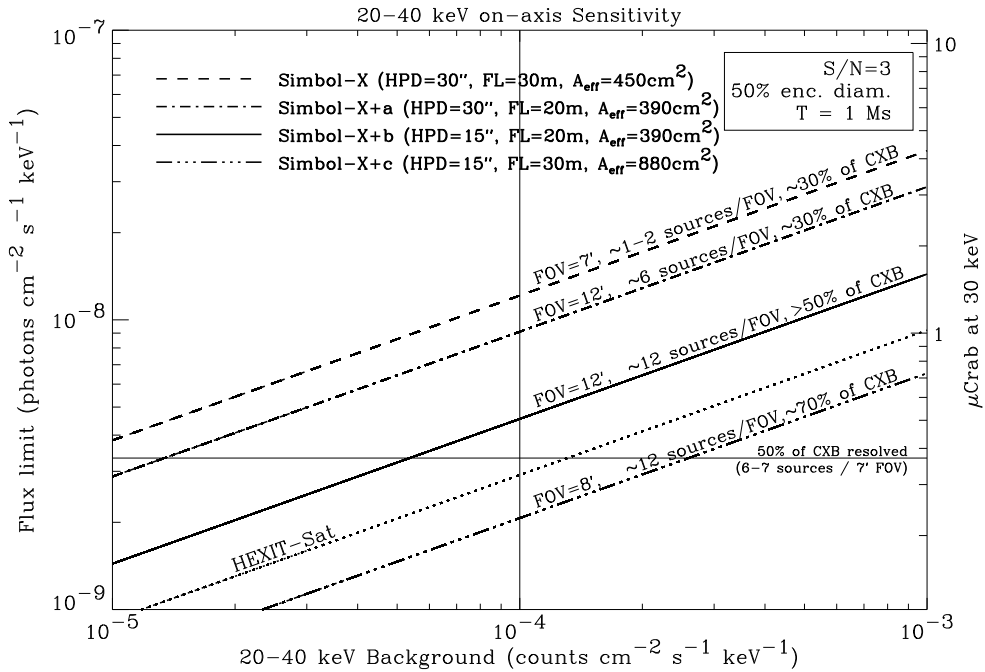}
   \end{tabular}
   \end{center}
   \caption[example] 
   { \label{fig:s_6} 
Expected continuum sensitivity in the 20--40 keV band for the current Simbol-X baseline hypothesis compared with the possible mission
enhancements (via multi-layer implementation, and mission parameter optmizzation), as a function of the detector background counting rate. The
reported number of sources refer to a $1\times10^4$ counts cm$^{-2}$ s$^{-1}$ keV$^{-1}$ background, and have been calculated taking into
account the increase of the telescope flux limit towards the FOV edge, and considering the effect of the different HPD values on the telescope
confusion limit. For comparison, the best telescopes at these energies, INTEGRAL/IBIS and BeppoSAX/PDS provided an angular resolution of 12$'$
(IBIS) and a sensitivity limit of $\sim300\mu$Crab (PDS).}
   \end{figure}

The value $A_{\rm d}$ depends upon the geometrical assembly of the telescope/detector and on the assumed fraction of focused photons, $\eta$. 
For $\eta=0.5$, then $A_{\rm d}$ is given by the following relation:
\begin{equation}
\label{eqn:Ad}
A_{\rm d} = \frac{\pi} {4} [\tan({\rm HPD}\cdot{\rm FL})]^2 \;\;\;\;\;{\rm cm}^{2}\;,
\end{equation}
where HPD is the Half Power Diameter angular resolution, and FL the Focal Length in cm.  Equations (\ref{eqn:Fmin}) and (\ref{eqn:Ad}) show
that since large focal lengths implicate large spot areas, with a consequent increase of the intrinsic background counts, the need to
optimize the telescope design in terms of sensitivity imply the necessity of trade-offs also in terms of FL and HPD. The latter, moreover,
having further importance in avoiding confusion limited deep observations. Figure \ref{fig:Ad_2} shows that an angular resolution around
$15''$ requires, for FL$=8\div10\;{\rm m}$, a spatial resolution below $300\mu{\rm m}$, which relax to $\sim400\div500\mu{\rm m}$ for
FL$=20\div30\;{\rm m}$, to sample the point spread function (PSF) with at least two pixels. On the other hand, the possibility to maintain a
moderate FL ($\sim20$m), by using multi-layer coated mirrors, allows for a field view (FOV) as wide as FOV=$15'$ (FHWM diameter). Such a
large FOV has fundamental scientific implications as discussed later in section 3.1 and shown in figure \ref{fig:s_6}.

Total reflection occurs for incidence angles below the critical angle, $\alpha_{\rm c}$, which is directly proportional to the square root
of the reflecting optics material $\rho$, and inversely proportional to the incoming photon energy, $E$. The effective area, $A_{\rm eff}$,
of an X-ray mirror telescope, based on Wolter I optics, is approximately given by:
\begin{equation}
\label{eqn:Aeff}
A_{\rm eff} \propto (1-\xi) \cdot {\rm FL}^2 \cdot \alpha^{2}_{\rm c} \cdot R^2 \;\;\;\;\;{\rm cm}^{2}\;,
\end{equation}
where $\alpha_{\rm c}$ is the reflection critical (i.e. maximum) angle, $R$ is the reflectivity of the mirrors, and $\xi$ is the telescope
area loss due to the vignetting caused by the spider arms or by the finite thickness of the mirror shell walls.  For energies $E>10\;{\rm
keV}$ the cut-off angle, $\alpha_{\rm c}$, becomes very small, thus implying very small effective areas for traditional (${\rm FL}<8$m)
telescopes. For example, even in the case of high density material optics like Platinum ($\rho=21.4\;{\rm g/cm}^{3}$), the value of
$\alpha_{\rm c}$ at 30 keV is more than five times smaller than at 5 keV (0.15$^\circ$ against 0.8$^\circ$).

Equation (\ref{eqn:Aeff}) shows that $A_{\rm eff}$ can be increased, with an associated sensitivity improvement, either
by having a large focal length, and/or a higher upper threshold value for $\alpha_{\rm c}$, which can be attained by using multi-layer
mirrors. Conversely, figure \ref{fig:Ad_2} shows that a longer FL implies a larger spot area (factor of $\sim10-15$ increase in $A_{\rm D}$, if 
going from FL=8m to FL=30m), and therefore a higher residual background counting rate.

\section{Background and sensitivity requirements: \\ the Simbol-X mission as a case study}

The contribution to the CXB due to highly absorbed (i.e.: Compton-thick) AGN, which are undetected in the canonical 2--10 keV band, is
$\approx50-70$\% of the total.  This means that in order to discover significant samples of this class of objects, we need to resolve into
discrete sources at least 50\% of the 20--40 keV CXB. Using the current AGN CXB synthesis models$^7$, this implies 
a limit sensitivity of $\sim7 \times 10^{-15}$ erg cm$^{-2}$ s$^{-1}$ or $\sim0.75\mu$Crab in the 20--40 keV band, coupled with an angular
resolution HPD$<15''\div20''$ in order to avoid, or minimize to $<20$\%, confusion problems$^1$.

The problem of extending soft X-ray capabilities, in terms of sensitivity and angular resolution, also at energies greater than 10--20 keV, was 
previously tackled for the HEXIT-Sat mission$^1$, which was specifically taylored for this study. This problem has then been addressed by the 
Simbol-X mission concept study as one of its key scientific objectives.  The Simbol-X mission$^5$ is an international collaboration led by the 
CEA/Saclay (France), with groups from Germany and Italy, aiming at a significant leap in the understanding of the sky above 10 keV.  The basic 
Simbol-X principle is to increase the effective area at E$>$10 keV by means of a long focal length (up to 30 m), while maintaining single-layer 
Wolter I mirror optics, and allowing maximum optics diameters of the same size as the ones used for XMM-Newton, i.e. $\approx$60-70 cm. The 
envisaged mission scenario is to place the detectors and the optics in two separate spacecrafts, exploiting the formation flight configuration 
concept to be implemented by CNES. The focal plane detector system combines a Si low energy detector, effective up to $\sim20$ keV, with a 
Cd(Zn)Te high energy detector. This hybrid solution allows wide band coverage ($\sim$0.1--70 keV), and spectral capabilities 
($\Delta$E$\simeq$120eV at 6 keV; $\Delta$E/E $<3$\% at 60 keV).

Simbol-X is now at the end of the Phase-0. The selection for the Phase-A is foreseen for autumn 2005, for a beginning of the Phase-A in January 
2006. A strong interest has been expressed by the italian astrophysics community, and by the Italian Space Agency (ASI), for a high level 
participation to the Simbol-X mission, provided a science-driven optimization of the mission design. The results presented here, specifically 
oriented to a scientific optimization of the current Simbol-X baseline hypothesis, can be applied, in principle, to any hard X-ray telescope based 
on focusing optics.

\subsection{Sensitivity requirements and evaluation}

In its current baseline hypothesis (single-layer optics, FL=30m, HPD=30$''$, FOV=7$'$), Simbol-X has a sensitivity (see figure \ref{fig:s_6})
$F_{\rm lim}\simeq1.4\mu$Crab (at 30 keV, in 1 Ms, 3$\sigma$, for a background of $1\times10^{-4}$cts cm$^{-2}$ s$^{-1}$ keV$^{-1}$). This
translates into just $\sim2$ sources/FOV, resolving about 35\% of the CXB at E$>$20 keV.  The feasibility studies conducted in 2003 for the
HEXIT-Sat mission concept$^1$ have shown that these performances can be enhanced by using multi-layer optics and by optmizing the telescope
design.  In figure \ref{fig:s_6} the minimum detectable continuum flux is shown as a function of the expected background, in the $20-40$ keV
energy band. The possible design optimizations choices shown in figure \ref{fig:s_6}, while remaining technically feasible within the Simbol-X
mass budget and construction constrains, can allow a sensibile improvement of its scientific performance. For the purpose of this study we have
focalized on the study of the CXB around 30 keV, and therefore on maximizing the fraction of resolved sources. This can be obtained in several
ways: adopting the telescope configuration parameters to reach a deeper sensitivity in a possibly greater FOV, optimizing the mirror shells in
order to have the best possible HPD to minimize confusion problems, or by combining the two options.

One of the most interesting mission options is the one labeled as {\em Simbol-X+b} in figure \ref{fig:s_6} (continuous line). The key parameters 
in this option are: FL=20m, HPD=15$''$, and $A_{\rm eff}$=390cm$^{2}$. The 30\% reduction of the focal length (from 30m to 20m, dot-dashed and 
continuous lines in figure \ref{fig:s_6}), together with the use of multi-layer mirrors, while causing on one side a slight decrease in effective 
area (see equation \ref{eqn:Aeff}), would allow a smaller spot area (see equation \ref{eqn:Ad}), and a factor $\sim$2 (in diameter) greater FOV. 
In the Simbol-X+b option, these factors are associated to a factor 2 improvement in HPD (from $30''$ to $15''$), and result in a factor $\sim$2.5 
improvement in sensitivity. Assuming a total detector background of $1\times10^4$ counts cm$^{-2}$ s$^{-1}$ keV$^{-1}$, this translates in the 
possibility to resolve into discrete sources more than 50\% of the CXB.

On the other hand, the use of the maximum diameter optics ($\sim70$cm) which can be hosted within the foreseen spacecraft assembly, coupled with 
shells capable to reach a $15''$ HPD, could indeed result feasible even in the presence of the difficulties posed by the high weight. With these 
optics, and maintaining the 30 m baseline FL, the effective area would be greater than $800$cm$^{-2}$, with a slightly wider FOV ($8'$ in 
diameter). This would imply a sensitivity of $\sim0.2\mu$Crab, which, at this FOV and HPD, would mean to resolve at least 70\% of the CXB sources. 
Moreover, it is important to point out that a further, independent, improvement in the CXB resolution, can be attained by having an even finer 
HPD, with a consequent minimization of the confusion problems which could arise at these deep sensitivity limits. As a comparison, even because of 
the absence of imaging capabilities in a $1^\circ$ FOV (FWHM), BeppoSAX/PDS had already very significant confusion problems with a factor 
$\sim50-100$ worse sensitivity.

   \begin{figure}
   \begin{center}
   \begin{tabular}{c}
   \includegraphics[scale=0.7]{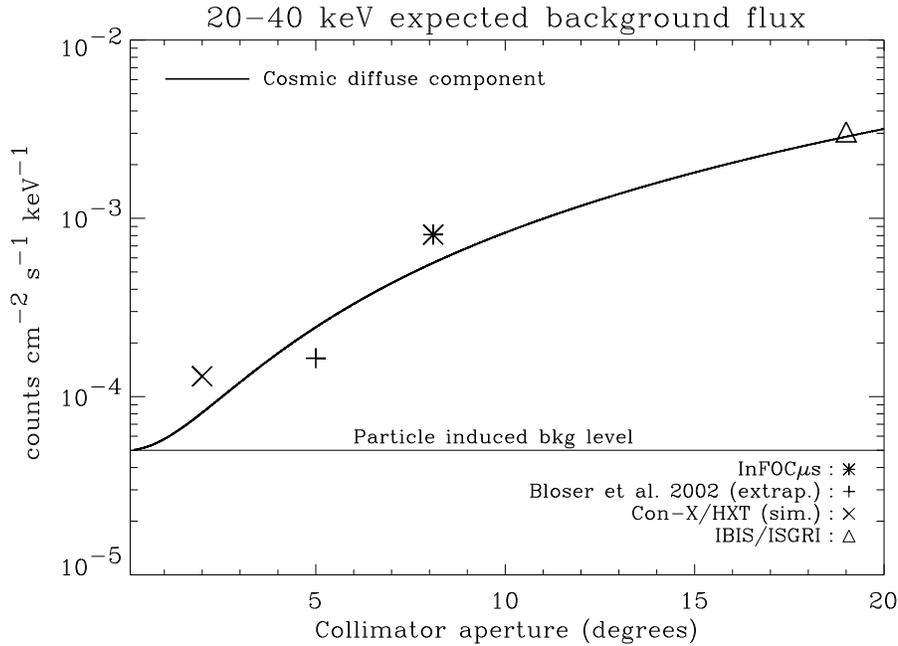}
   \end{tabular}
   \end{center}
   \caption[example] 
   { \label{fig:coll_1} 
Expected background flux in the 20-40 keV band as a function of the collimator aperture. The continuous line 
represents the diffuse CXB covolved with the collimator response, while the thin horizontal line shows the particle
induced background level re-normalized from BeppoSAX/PDS. The Bloser et al. (2002) point is an extrapolation from a 
measurement performed with a $\sim1600$cm$^2$ aperture, while the Con-X/HXT point refers to a simulation results (see 
text). Already at narrow aperture angles, the expected/measured spectrum is dominated by the diffuse component is 
expected to dominate, and the requirement to have a background $\leq1\times10^{-4}$ counts cm$^{-2}$ s$^{-1}$ 
keV$^{-1}$, poses a limit of $\simeq3^\circ$ (in diameter) to the collimator aperture angle.}
   \end{figure} 

\subsection{Background evaluation and active/passive shielding requirements}

The background in high energy telescopes is composed by two main components: one is hadron induced, while the second one is directly due to the 
high energy photons of the cosmic X-ray background. The absolute magnitude of the two components, and the ratio between the two, vary with energy, 
and depend in principle upon the telescope geometrical set-up (detector material and thickness, field of view, passive shielding design and 
materials, etc...), the background minimization philosophy (passive and active shielding design, anticoincidence threshold, {\em good} events 
characterization and reconstruction, etc....), and the telescope mission scenarios (orbit choice, spacecraft configuration, ...).

The popularity of CdZnTe (CZT) as focal plane detector material for hard X-ray focusing telescopes has constantly increased in recent years, as it 
has been proposed for several high energy mission projects or ideas: HEFT$^8$, NuSTAR$^9$, Constellation-X/HXT$^{10}$, XEUS$^6$, 
InFOC$\mu$S$^{11}$, Simbol-X$^5$, HEXIT-SAT$^1$. The commonality among the current designs is the placement of the CZT inside deep (active and/or 
passive)  anticoincidence wells. However, these types of instruments have still to considered to be in their pioneering phase, since on-orbit 
measurements, performed with focussing CZT telescopes above 10-20 keV, are still lacking.

Recent balloon experiments$^{11,12}$, based on an actively (organic plus inorganic) and passively shielded CZT detectors, have measured a 
background level which varies, depending on the detector configuration and aperture angle. These measurements are reported in figure 
\ref{fig:coll_1}, superimposed to the expected CXB flux contribution$^{13}$, as a function of collimator aperture. The InFOC$\mu$S 
measurement$^{11}$ refers to the July 2001 flight performed with a 2mm thick CZT detector which had a shield opening angle of $8.1^\circ$. The 
other balloon result$^{12}$ shown in figure \ref{fig:coll_1} is instead the extrapolation down to a $5^\circ$ aperture of a measurement performed 
with a 2mm thick CZT detector having a very wide aperture angle ($\sim40^\circ$). These two balloon measurements have been normalized to a cosmic 
environment, assuming a factor $\sim3$ reduction due to residual (3 g/cm$^{-2}$ at balloon altitudes) atmospheric absorption. The point at 
$2^\circ$ aperture is the result of the MonteCarlo evaluation performed for the Constellation-X/HXT instrument in the case of a L2 orbit 
scenario$^{14}$. Finally, the point at $19^\circ$ corresponds to the background measured by the 2mm thick CdTe ISGRI detector onboard the IBIS 
telescope$^{3}$.

The PDS detector$^2$ onboard the BeppoSAX satellite, based on NaI-CsI phoswich with passive ($\sim1.3^\circ$) and active (CsI+plastic) shields, 
has measured a particle induced background level which, scaled to a 2 mm detector thickness, is equal to $\sim5\times10^{-5}$ cts cm$^{-2}$ 
s$^{-1}$ keV$^{-1}$. This value, shown as the horizontal line in figure \ref{fig:coll_1}, can be used as a reference for the hadronic component, 
to which one has to sum the contribution of the CXB component to get the total expected background.

Figure \ref{fig:coll_1} shows that already for very small collimator aperture angles, the measured/expected background at 20--40 keV is dominated 
by the diffuse component. In fact, the requirement to keep the background $\leq10^{-4}$ counts cm$^{-2}$ s$^{-1}$ keV$^{-1}$ implies a collimator 
aperture angle $<3^\circ$ (in diameter). It is important to remind that for this evaluation we have assumed to reject all the possible 
fluorescence line photons contributions from the passive shielding material itself. This issue will be addressed in section 4, where the possible 
technical implementations are presented and discussed. The background evaluation shown in Figure \ref{fig:coll_1} has been scaled from balloon and 
satellite experiments in which the CZT detector was actively and passively shielded. The passive shielding was done by means of a graded 
(Pb-Sn-Cu) collimator. The active shielding included either a BGO (Constellation-X/HXT, IBIS/ISGRI) or a CsI (InFOC$\mu$S) on the rear of the 
detection plane, while only in one case$^{12}$ the possibility to use only plastic has been taken into consideration.

As pointed out in the Constellation-X/HXT simulation report$^{14}$, a thick active veto based upon inorganic detector could determine an 
"overshielding" of the particle induced component, with a consequent increase of the secondary (via isotope activation) emission.  For the project 
under study, however, the passive and active shield design can follow a slightly different philosophy. In fact, given the low, $\sim80-100$ keV, 
upper energy threshold, an active/passive shielding system based upon plastic (or inorganic scintillator, but with a thickness significantly 
smaller than what used for detectors operating in the hundreds of keV region) coupled with a graded collimator can be feasible. Thus, a thinner 
BGO (or CsI), or a complete replacement of the inorganic crystal with plastic would significantly decrease the activation components, due to 
radioactive isotopes, which are known to contaminate the background spectrum in the tens of keV region.

\subsection{Telescope design concepts}

\begin{figure}
   \begin{center}
   \begin{tabular}{c}
   \includegraphics[scale=0.70]{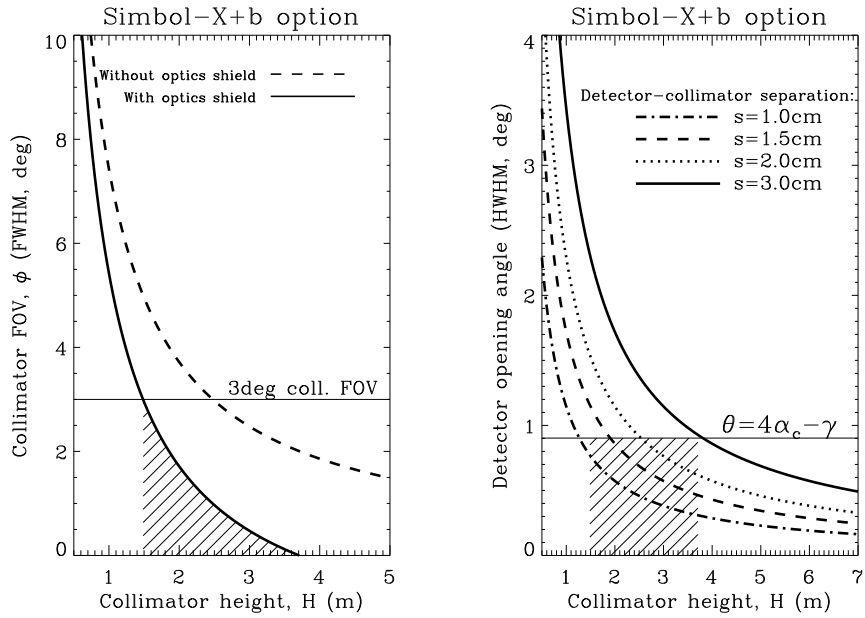}
   \end{tabular}
   \end{center}
   \caption{ \label{fig:H_2}
Analysis of the possible trade-offs between collimator FOV ($\phi$), collimator height ($H$), detector opening angle, detector-collimator 
separation ($s$), and maximum photon incidence angle on the detector ($4\alpha_{\rm c}-\gamma$, see text and equation 4), for the Simbol-X+b 
(FL=20m, FOV=$12'$) mission concept.  {\em Left panel:} Collimator acceptance angle (FWHM FOV) as a function of collimator wall height. The two 
curves indicate the collimator aperture, with (continuous curve), and without (dashed) taking into account the screening contribution of the 
optics structure/spacecraft (in the conservative hypothesis of no circular baffling around the optics).  {\em Right panel:} Detector opening angle 
as a function of collimator walls height, for different detector-collimator separation values ($s$). Superimposed is the horizontal line 
corresponding to the maximum photon incidence angle on the detector. The shaded areas in both panels indicate the parameter space allowed in order 
to have a collimator FOV $\phi=\le3^\circ$ (left panel), while avoiding vignetting of the focused photons by the collimator walls (right panel).  
}
   \end{figure}

\begin{figure}
   \includegraphics[scale=0.8,clip,trim=0 150 0 60]{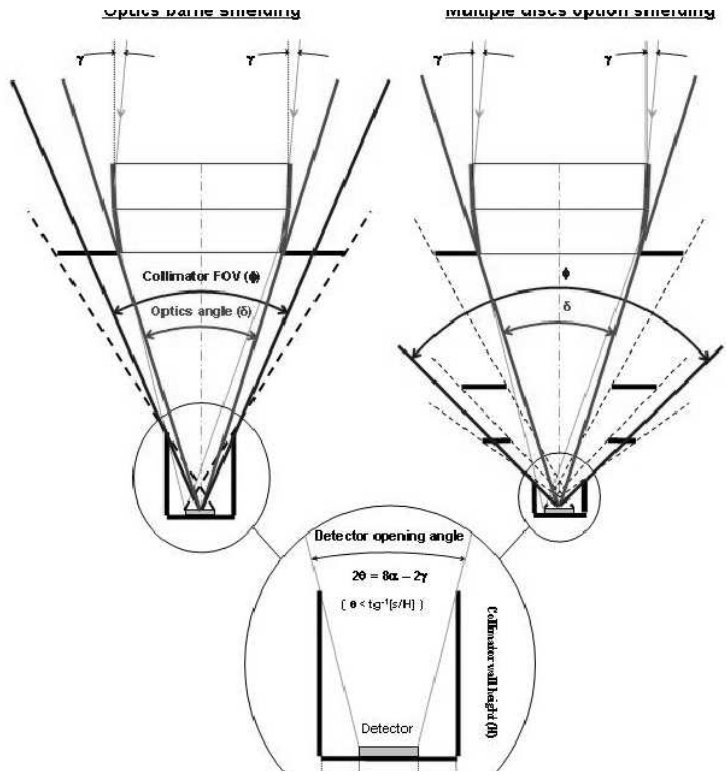}
   \caption{ \label{fig:sb} 
Detector-optics passive shielding concepts: $\gamma$ = maximum incidence angle for the most external shell; $\phi$ = collimator FOV (FWHM);  
$\delta$ = angle subtended by the optics at the detector. {\em Top left panel:} Simple collimator plus optics baffling. {\em Top right panel:} 
Shorter collimator coupled with multiple discs structure plus optics baffling. {\em Bottom panel:} In order to avoid vignetting by the collimator 
walls, the choice of $s$ (detector-collimator separation) and $H$ (detector walls height) must allow a detector opening angle greater than the 
maximum focused photon incidence angle on the detector. }
   \end{figure}

Figure \ref{fig:coll_1} shows that the photon diffused background component dominates above a few degrees collimator
aperture, and that in order to meet the required limit of a $1\times10^{-4}$ counts cm$^{-2}$ s$^{-1}$ keV$^{-1}$
background, the collimator aperture must be smaller than $2^\circ-3^\circ$. On the other hand, given the structure of
the focal plane detector and the further shielding due to the optics, the FOV defined by the collimator is not the only
parameter to determine the aperture of the focal plane detector to the diffuse background. In fact, in a simple 
well-type configuration (see also figure \ref{fig:sb}), the height, $H$, of the collimator walls, necessary to reach a limiting
collimator FOV is given by:
\begin{equation}
\label{eqn:H}
H=\frac{FL \cdot \tan\left({\frac{\textstyle\gamma}{\textstyle 2}}\right)+s} 
{\tan \left({\frac{\textstyle\phi}{\textstyle 2}+\frac{\textstyle\delta}{\textstyle 2}}\right)},
\end{equation}
where:\\
\begin{tabular}{ll}
$\phi$:           & collimator FOV (FWHM) limit for background minimization; \\
$s$:              & separation between detector and collimator walls; \\
$\delta$:         & angle subtended by the optics structure at the detector; \\
$\gamma$:         & maximum incidence angle for the most external shell; \\
$\alpha_{\rm c}$: & optics critical angle. \\
\end{tabular}\\
The photons which impinge on the most external shell with an angle $\gamma$, will be reflected towards the detector with an angle 
$\theta=4\alpha_{\rm c}-\gamma$. 
Therefore, in order to have a detector {\em opening angle} large enough to avoid the possible vignetting caused by the collimator walls, 
the value of $s$ must be $ s > H\tan{(4\alpha_{\rm c}-\gamma)}$ (see figure \ref{fig:H_2} and figure \ref{fig:sb}).

\begin{table}
\caption{\label{tab:fov} Design concept key parameters for having a maximum collimator aperture of 3$^\circ$ (corresponding to $\approx 5\div10 
\times 10^{-5}$ cts cm$^{-2}$ s$^{-1}$ keV$^{-1}$ in the 20-40 keV band. The evaluations refer to the Simbol-X mission (baseline hypothesis, and 
Simbol-X+b option), and HEXIT-Sat. A simple geometry has been assumed, taking into account only the collimator walls and the covering determined 
by the optics structure. A detector-collimator separation $s=3$cm has been used (see text and figure \ref{fig:H_2}).}
\begin{tabular}
{   l          |            c          c                       cc                             cc}             \hline
Telescope      & FOV             & Focal Length & \multicolumn{2}{c}{Mirror optics} & \multicolumn{2}{c}{Collimator height (cm)} \\ 
               &  (FWHM)         & (m)          & Diam. (cm) & Subt. angle (deg)    & 3$^\circ$ apert. & compl. shield  \\ \hline
Simbol-X       &  $7'$           & 30           & 60         & 1.2                  & 167.3            & 605.4     \\
Simbol-X+b     &  $12'$          & 20           & 70         & 2.0                  & 148.5            & 370.9     \\
HEXIT-Sat      &  $17'$          & 8            & 30         & 2.2                  & 110.7            & 265.5     \\
\hline
\end{tabular}
\end{table}
In order to quantify the requirements in terms of collimator walls height, imposed by having a narrow detector aperture, we have taken the Simbol-X
mission as a case study, and compared its current baseline hypothesis with an improvement option, which is referred as ``Simbol-X+b'' in figure
\ref{fig:s_6}. The results are shown in table \ref{tab:fov}.  For this calculation we have
assumed a simple geometry, taking into account only the collimator walls and the covering determined by the optics structure. The possibility to
place a circular shielding structure around the optics would result in a further reduction of the collimator angle, or, for a given detector
aperture, in a shorter collimator walls height (see section \ref{sec:opt}). In the case of Simbol-X the requirement of a $3^\circ$ collimator
aperture translates into a collimator wall height of about 1.7m, slightly higher than what needed for HEXIT-Sat because of its shorter FL.
In the Simbol-X+b option, the shorter FL allows for significantly shorter collimator walls also in the case of an even narrower collimator aperture.

The values reported in table \ref{tab:fov} have been calculated for a detector-collimator separation, $s=3$cm. A non-zero value of $s$ is 
necessary to avoid vignetting caused by the collimator itself. Indeed, if, on one side, the detector must be shielded against unwanted background 
photons, the shielding assembly, on the other hand, must not stop photons coming from within the telescope field of view. This can be obtained by 
separating the collimator walls from the active detector by a distance, $s$. In this way, the angular response of the detector will have the 
canonical shape of a collimated instrument, but with a plateau in the centre. By accurately choosing the combination between collimator height and 
detector-walls separation, the angular dimension of this plateau (i.e. the detector opening angle) will avoid the vignetting of the focussed 
X-rays. Figure \ref{fig:H_2} shows an example of the trade-off analysis between the three relevant parameters: collimator aperture angle, 
collimator walls height, and detector-collimator separation for the Simbol-X+b (FL=20m, FOV=$12'$) mission concept.

The left panel shows the collimator acceptance angle (FWHM FOV) as a function of collimator wall height. The two curves indicate the collimator 
aperture with (continuous curve) and without (dashed) subtracting the angle subtended at the detector by the optics structure/spacecraft (in the 
conservative hypothesis of no circular baffling around the optics).  The right panel of figure \ref{fig:H_2} shows the detector opening opening 
angle as a function of collimator wall height, for different detector-collimator separation values ($s$). Superimposed is the horizontal line 
corresponding to the maximum photon incidence angle on the detector. The shaded areas in both panels indicate the parameter space allowed in order 
to have a collimator FOV $\phi=\le3^\circ$ (left panel), while avoiding vignetting of the focused photons by the collimator walls (right panel). 
Even in the case of a complete shielding scenario, in the Simbol-X+b configuration the vignetting would be avoided by having a collimator-detector 
distance $s=3$cm.

\section{Design optimization and implementation studies}
\label{sec:opt}

   \begin{figure}
   \begin{center}
   \begin{tabular}{c}
   \includegraphics[scale=0.77]{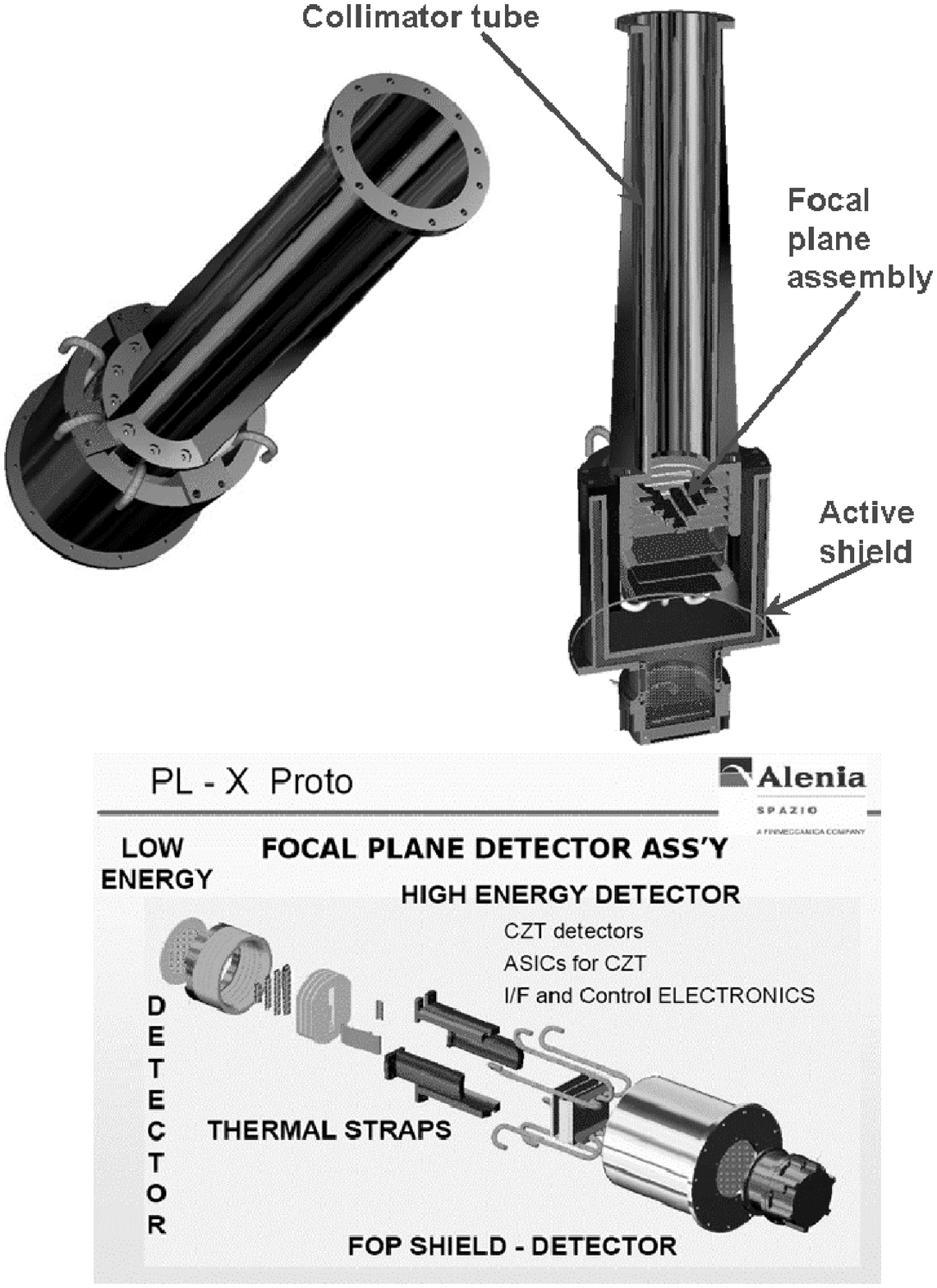}
   \end{tabular}
   \end{center}
   \caption{ \label{fig:jmp} 
Different views of one of the possible technical implementations which resulted from the pre-feasibility study performed
by LABEN/Alenia-Spazio within the framework of a project financed by the Italian Space Agency (ASI). The exploded view
({\em right} and {\em bottom} panel) shows the various subsystems: collimator, focal plane assembly, rear active 
shield, together with the thermal and electronics interfaces.}
   \end{figure} 

The scientific requirement of having a narrow field of view for the focal plane detector can be met by means of a number of possible
configuration designs for the shielding assembly. The technical approach can be divided into two separate solution philosophies which are shown
in figure \ref{fig:sb}: (a) A simple tube-shaped shield (top left panel of figure \ref{fig:sb}); or (b) a shorter, tube-shaped, passive shield,
coupled with concentric and coplanar discs, placed at different heights above the detector (top right panel). The zoomed inset of figure
\ref{fig:sb} clearly shows how the necessity (which has been quantified in details in figure \ref{fig:H_2} and table \ref{tab:fov}) to avoid
vignetting by the collimator implies a careful trade-off analysis between $H$ (collimator height), $s$ (detector-collimator separation),
$\delta$ (the angle subtended by the optics at the detector, $\alpha_{\rm C}$, and the telescope FOV.  Both options can then be completed with
the addition of a disc-shaped structure around the optics (or optics spacecraft, in the formation flight scenario) to achieve a further
shielding without increasing the height of the collimator walls.

Solution (a) is naturally feasible for any telescope/detector configuration. Solution (b), on the other hand, can be implemented in 
the HEXIT-Sat scenario, using the telescope structure itself, while for a formation flight configuration, the use of an extendible bench, in a 
concept similar to what proposed for sun shielding of the XEUS gratings$^{15}$ (in case of their implementation), is feasible.

Within the activity of a pre-feasibility study performed with LABEN/Alenia-Spazio, in the framework of a project financed by ASI, a detailed 
technical study for a hard X-rays telescope meeting the scientific requirements outlined in section 1, has been
carried out.  The key aims of the project activity have been the following: (a) definition of the general configuration of the proposed
telescope (detection plane, plus active and passive shielding systems); (b) mechanical and thermal design and assessment study of the focal
plane; (c) passive shield feasibility study for different possible configurations, with associated design trade-offs and detailed analysis.
Figure \ref{fig:jmp} shows one of the possible technical implementations which resulted from this study. The envisaged telescope concept
included: a low energy ($\sim0.1-10$ keV) detector possibly based on Silicon drift detectors; a high energy detector ($\sim5-70$ keV)
based on Cd(Zn)Te room-temperature crystals coupled to a dedicated front end electronic ASIC, a rear active shield, and a tube-shaped
collimator structure. In addition to the requirement of having a low background, the main design drivers have been:  to minimize the length of
the thermal paths from the CZT/ASIC assembly to the spacecraft thermal interface; to allow a good distributions of CZT/ASIC power
dissipation ($2\div3$ W per detector layer), in order to have a uniform temperature across the detection plane; to ensure a modularity design
of the high energy detector, to allow for possible non-destructive reworking or detector refurbishment.

Recent measurements$^{12}$ have indicated that a well taylored combination of passive and plastic shielding can offer an acceptable background level
in tens of keV region. The possibility to avoid inorganic crystals (e.g. CsI, BGO) would have several advantages: plastic is lighter, can be reworked
into dedicated shapes without requiring the use of separated ingots, and, most importantly from a scientific point of view, it does not create the
activation lines due to radioactive nuclides. In this framework, we have studied the feasibility (see figure \ref{fig:jmp}) of an active shield
composed by plastic (e.g. BG--408), surrounded by a high-Z (Tantalum Z=73, or Tungsten Z=74) graded passive envelope.

Great effort has been dedicated in studying the graded shield configuration in order to minimize the residual background due to K$\alpha$ line
fluorescence emission. Four alternative combinations of graded shields have been considered using Tin (Z=50) as the first layer, and Copper,
Aluminium, and Carbon as additional layer. The use of Carbon as the external layer would minimize the emission also at $E<8$ keV in order not to
contaminate the low energy detector. The thickness of the layers, $T_{\rm L}$ has been chosen to satisfy the relation $T_{\rm L} > \mu(E_{\rm
F})^{-1}$, where $\mu(E_{\rm F})$ is the attenuation coefficient at the highest fluorescent energy, $E_{\rm F}$, of the neighbouring layer. The total
thickness of the various configurations is always below 1.4mm, also in the case of using Carbon which, having a low $\mu$ requires the largest
thickness.

Further studies are programmed in order to realize, test and calibrate prototypes of both the focal plane detector assembly and the multi-layer
optics. In parallel, a dedicated activity is ongoing in order to identify the critical telescope design issues to be addressed to improve the current
mission baseline hypotheses.

\section{Conclusions}

A detailed study of the sensitivity for a hard X-ray focussing telescope has been carried out. In addition to the already known requirements of a 
low particle induced background and a fine HPD to minimize confusion problems, the importance to shield against the diffuse background has 
emerged. The background diffuse component is expected to dominate for aperture angles greater than a few degrees, and the sensitivity requirements 
impose a maximum collimator aperture of $3^\circ$ (diameter, FWHM), in order to maintain the background below $1\times10^{-4}$ counts cm$^{-2}$ 
s$^{-1}$ keV$^{-1}$ at $\approx$30 keV.

The Simbol-X mission has been taken as a case study, and different options, still compatible with the mission mass and dimension budget constraints,
have been envisaged which would allow an improvement of the scientific performance with respect to the present baseline hypothesis, having as a key
science objective the study of the CXB in the region where it peaks ($20-40$ keV). The improvements can be attained mainly by decreasing the FL (30m
to 20m) and improving the HPD ($30''$ to $15''$), with a simultaneous increase of the FOV ($7'$ to $12'$) by means of multi-layer implementation.

The possible telescope-collimator-detector configurations have been investigated in terms of parameters optimization. The collimator height 
necessary to maintain the detector aperture below $3^\circ$ has been calculated taking into account the additional shielding offered by the 
optics. A collimator height $H\sim1.5$m would ensure the required $3^\circ$ aperture. A narrower collimator aperture would need either higher 
collimator walls, or a disc-shaped shield structure around the optics to increase the covering angle.

A trade-off study between collimator walls height and detector-collimator separation has been carried out and the feasibility to avoid vignetting 
has been demonstrated also for the smallest (i.e. complete shielding) collimator aperture angles, by means of an acceptable collimator-detector 
separation $s\simeq3$cm.

The result of a feasibility study of a focal plane assembly coupled to an active and passive shielding design concept has
been presented.

\acknowledgments     
This work has been financed by the Italian Space Agency (contract I/014/04/0). 
The use of the facilities of LABEN/Alenia-Spazio is kindly acknowledged.
We thank Prof. O. Citterio and Prof G.C. Perola for extensive and extremely fruitful discussions.

\section{REFERENCES}

\begin{enumerate}

\item Fiore, F., et al. 2004, {\em HEXIT-Sat: a mission concept for X-ray grazing incidence telescopes from 0.5 to 70 keV}, Proc. SPIE, 
Vol. 5488, p. 933-943

\item Frontera F., et al 1997, {\em The high energy instrument PDS on-board the BeppoSAX X--ray astronomy satellite}, A\&A Supplement
series, Vol. 122, p. 357-369

\item Ubertini, P., et al. 2003, {\em IBIS: The Imager on-board INTEGRAL}, A\&A, Vol. 411, p. L131-L139

\item Ramsey, B., et al.  2002, {\em First images from HERO, a hard X-ray focusing telescope}, ApJ, Vol. 568, p. 432-435

\item Ferrando, P., et al. 2004, {\em SIMBOL-X: a new-generation hard x-ray telescope}, Proc. SPIE, Vol. 5168, p. 65-76

\item Parmar, A., et al. 2004, {\em Science with XEUS: the X-Ray Evolving Universe Spectroscopy mission}, Proc. SPIE, Vol. 5488, p.
388-393

\item Comastri et al. 1995, {\em The contribution of AGNs to the X-ray background}, A\&A, Vol. 296, p. 1

\item Harrison, F., et al. 2004, {\em Development of the High-Energy Focusing Telescope (HEFT) Balloon Experiment}, Proc. SPIE, Vol.
4012, p. 693-699

\item Harrison, F. 2004, {\em Nuclear Spectroscopic Telescope Array (NuSTAR) mission: Imaging the Hard X-ray Sky}, American
Astronomical Society, HEAD meeting \#8, \#41.05

\item Tumer, T., et al. 2004, {\em Test results of preliminary CdZnTe pixel detectors for possible application to HXT on the
Constellation-X mission}, Proc. SPIE, Vol. 5165, p. 548-553

\item Baumgartner, W. H., et al. 2003, {\em InFOCuS hard x-ray telescope: pixellated CZT detector/shield performance and flight
results}, Proc.  SPIE, Vol. 4851, p. 945-956

\item Bloser, P.F., et al. 2002, {\em Balloon flight background measurement with actively-shielded planar and imaging CZT detectors},
Proc. SPIE Vol. 4497, p. 88-99

\item Zombeck, M. 1991, {\em Handbook of space astronomy and astrophysics}, 2nd ed., CUP, p. 197

\item Armstrong, T.W., et al. 1999, {\em Initial estimates of radiation backgrounds for the hard X-ray telescope (HXT) on the planned Constellation-X 
mission}, Report No. SAIC-TN-99015R3

\item Gerlach, L., et al. 2004, {\em XEUS Concurrent Design Facility}, ESA, CDF--31(A)

\end{enumerate}

\end{document}